\documentclass{aa}
\usepackage[acronym,toc,nonumberlist]{glossaries}

\newacronym{CNES}{CNES}{Centre National d'Etudes Spatiales}
\newacronym{DSN}{DSN}{Deep Space Network}
\newacronym{ESA}{ESA}{European Space Agency}
\newacronym{GINS}{GINS}{G\'eod\'esie par Int\'egrations Num\'eriques Simultan\'ees}
\newacronym{INPOP}{INPOP}{Int\'egrateur Num\'erique Plan\'etaire de l'Observatoire de Paris}
\newacronym{LOS}{LOS}{line of sight}
\newacronym{NAIF}{NAIF}{Navigation and Ancillary Information Facility}
\newacronym{NASA}{NASA}{National Aeronautics and Space Administration}
\newacronym{PDS}{PDS}{Planetary Data System}
\newacronym{SEP}{SEP}{Sun Earth Probe}
\newacronym{VEX}{VEX}{Venus Express}
\newacronym{MGR}{MESSENGER}{MErcury Surface, Space ENvironment, GEochemistry, and Ranging}
\newacronym{JPL}{JPL}{Jet Propulsion Laboratory} 
\newacronym{RMDCT}{RMDCT}{Radio Metric Data Conditioning Team}
\newacronym{ODF}{ODF}{Orbit Data File}
\newacronym{EOP}{EOP}{Earth Orientation Parameters}
\newacronym{IAU}{IAU}{International Astronomical Union}
\newacronym{HEF}{HEF}{mount high efficiency}
\newacronym{BVE}{BVE}{block 5 exciter}
\newacronym{COI}{COI}{center of integration}
\newacronym{PPN}{PPN}{Parametrized Post Newtonian}
\newacronym{ROB}{ROB}{Royal Observatory of Belgium}
\newacronym{rms}{rms}{root mean square}
\newacronym{GR}{GR}{general relativity}

\makeglossaries

\usepackage{graphicx}
\usepackage{natbib}
\usepackage[utf8]{inputenc}
\usepackage{graphics}
\usepackage[subnum]{cases}
\usepackage{threeparttable}
\usepackage[titletoc,title]{appendix}
\usepackage[dotinlabels]{titletoc}
\usepackage{makecell}
\usepackage[colorlinks=true,linkcolor=red,citecolor=blue]{hyperref}
\usepackage{lscape}
\usepackage[figuresright]{rotating}
\usepackage{multirow,booktabs}
\usepackage{ulem}
\usepackage{txfonts}
\usepackage{colortbl}                          
\usepackage{stfloats}
\usepackage{caption}
\usepackage{subcaption}
\usepackage{enumerate}
\usepackage[b]{esvect }
\usepackage{textcomp}

\begin{document}

\title{Use of MESSENGER radioscience data to improve planetary ephemeris and to test general relativity. }

 \author{A. K. Verma\inst{1,2} \and A. Fienga\inst{3,4} \and J. Laskar\inst{4} \and H. Manche\inst{4} \and M. Gastineau\inst{4}
}

 \institute{ Observatoire de Besan\c con, UTINAM-CNRS UMR6213,
 41bis avenue de l'Observatoire, 25000 Besan\c con, France
  \and
 Centre National d'Etudes Spatiales, 18 avenue Edouard Belin, 31400 Toulouse, France
 \and
 Observatoire de la Cote d\textquotesingle Azur, G\'eoAzur-CNRS UMR7329, 250 avenue Albert Einstein, 06560 Valbonne, France
 \and
 Astronomie et Syst\`emes Dynamiques,
 IMCCE-CNRS UMR8028,
 77 Av. Denfert-Rochereau, 75014 Paris, France
 }

 \offprints{A. Verma, ashok@obs-besancon.fr}

 \date{\today}

 \titlerunning{Planetary ephemeris construction and test of relativity with MESSENGER}
 \authorrunning{Verma et al.}

  \abstract{The current knowledge of Mercury orbit has mainly been gained by direct radar ranging obtained from the 60s to 1998 and by five Mercury flybys made by Mariner 10 in the 70s, and MESSENGER made in 2008 and 2009. On March 18, 2011, MESSENGER became the first spacecraft to orbit Mercury. The radioscience observations acquired during the orbital phase of MESSENGER drastically improved our knowledge of the orbit of Mercury. An accurate MESSENGER orbit is obtained by fitting one-and-half years of tracking data using GINS orbit determination software. The systematic error in the Earth-Mercury geometric positions, also called range bias, obtained from GINS are then used to fit the INPOP dynamical modeling of the planet motions. An improved ephemeris of the planets is then obtained, INPOP13a, and used to perform general relativity tests of PPN-formalism. Our estimations of PPN parameters ($\gamma$ and $\beta$) are more stringent than previous results.

  \keywords{messenger - celestial mechanics - ephemerides - general relativity }}
 
  \maketitle


\section{Introduction}
Mercury is the smallest and least explored terrestrial planet of the solar system. Mariner 10 was the first spacecraft to make three close encounters (two in 1974 and one in 1975) to this mysterious planet, and it provided most of our current knowledge of the planet until early 2008 \citep{Smith10}. In addition to Mariner 10 flyby observations, ground-based radar measurements were the only observations to be used to study Mercury\textquoteright s gravity field and its physical structure (spherical body with slight flattening at the poles and a mildly elongated equator) \citep{Anderson87,Anderson1996}. In 2004, the \gls{NASA} launched a dedicated mission, \gls{MGR}, to learn more about this planet. \gls{MGR} made three close encounters (two in 2008 and one in 2009) to Mercury and became the first spacecraft to observe Mercury from its orbit.   

Untill now, \gls{MGR} has completed more than two years on orbit at Mercury. During the orbital period, radio tracking of \gls{MGR} routinely measured the Doppler and range observables at \gls{DSN} stations. These observables are important for estimating the spacecraft state vectors (position and velocity) and improving the knowledge of Mercury\textquoteright s gravity field and its geophysical properties \citep{Srinivasan07}. Using the first six months of radioscience data during the orbital period, \cite{Smith12} computed the gravity field and gave better constraints on the internal structure (density distribution) of Mercury. This updated gravity field becomes crucial for the present computation of \gls{MGR}  orbit and for performing precise relativistic tests.    

The primary objectives of this work are to determine the precise orbit of the \gls{MGR} spacecraft around Mercury using radioscience data and then to improve the planetary ephemeris INPOP \citep{Fienga2008,Fienga2009,Fienga2011}. The updated spacecraft and planetary ephemerides are then used to perform sensitive relativistic tests of the \gls{PPN} formalism \citep{Will93,Will01,Will06}. 

Nowadays, spacecraft range measurements are the most accurate measurements used for constructing planetary ephemerides. These measurements cover approximately 56$\%$ of all INPOP data \citep{Fienga2011} and impose strong constraints on the planet orbits and on the other solar system parameters, including asteroid masses. However, until now, only five flybys (two from Mariner 10 and three from \gls{MGR}) range measurements have been available for imposing strong constraints to Mercury\textquoteright s orbit \citep{Fienga2011}. Therefore, range measurements obtained by \gls{MGR} spacecraft during its mapping period are important for improving our knowledge of Mercury\textquoteright s orbit.  

Moreover, high-precision radioscience observations also offered an opportunity to perform sensitive relativistic tests by estimating possible violation of the two relativistic parameters ($\gamma$ and $\beta$) of the \gls{PPN} formalism of \gls{GR} \citep{Will93}. The previous estimations of these parameters using different techniques and a different data set, can be found in \cite{Bertotti03, Muller08, Pitjeva09, Williams09, Manche2010, Konopliv11, Fienga2011}. However, because of Mercury's relatively high eccentricity and its close proximity to the Sun, its orbital motion provides one of the best solar system tests of \gls{GR} \citep{Anderson97}. In addition, \cite{Fienga2010,Fienga2011} also demonstrated, Mercury observations are far more sensitive to \gls{PPN} modification of \gls{GR} than other data used in the planetary ephemerides. We, therefore, also performed the test of \gls{GR} with the latest \gls{MGR} observations to obtain one of the most precise value for \gls{PPN} parameters.

In this paper, we introduce the updated planetary ephemeris INPOP13a and summarize the technique used for estimating the \gls{PPN} parameters. The outline of the paper is as follows Section \ref{messenger} discusses the radioscience data analysis of the \gls{MGR} spacecraft. The dynamic modeling of \gls{MGR} and the results obtained during orbit computation are also discussed in the same section. In section \ref{inpop}, we discuss the construction of INPOP13a using the results obtained in section \ref{messenger}. In section \ref{test}, we discuss the gravitational tests using updated \gls{MGR} and Mercury ephemerides. Section \ref{dis} follows with conclusions and perspectives.

\section{MESSENGER data analysis}
\label{messenger}
\label{data}
Under NASA\textquoteright s Discovery program, the \gls{MGR} spacecraft is the first probe to orbit the planet Mercury. It was launched in August 3, 2004, from Pad B of Space Launch Complex 17 at Cape Canaveral Air Force Station, Florida, aboard a three-stage Boeing Delta II rocket. On March 18, 2011, \gls{MGR} successfully entered Mercury\textquoteright s orbit after completing three flybys of Mercury following two flybys of Venus and one of Earth \citep{Solomon07}.

The \gls{MGR} spacecraft was initially inserted into a $\sim$12-hour, near-polar orbit around Mercury, with an initial periapsis altitude of 200 km, initial periapsis latitude of 60\textdegree N, and apoapsis at $\sim$15,200 km altitude in the southern hemisphere. After a successful first-year flight in this orbit, the mission was extended to one or more years which began on March 18 2012. During first extended mission, two orbit-correction maneuvers were executed, four days apart, in April 2012 to reduce MESSENGER\textquoteright s orbital period from $\sim$12 to $\sim$8 hours \citep{Flanigan13}

The \gls{MGR} spacecraft was tracked by NASA\textquoteright s \gls{DSN} stations at X-band frequency, 7.2 GHz for a uplink from the ground stations and 8.4 GHz for a downlink from the spacecraft. Communications were accomplished via the 34-m and 70-m antennas of \gls{DSN} stations in Goldstone, CA; Madrid, Spain; and Canberra, Australia. MESSENGER\textquoteright s X-band tracking consists in measuring the round-trip time delay (two-way range) and the two- and three-way ramped Doppler shift of the carrier frequency of the radio link between the spacecraft and the \gls{DSN} stations on Earth. The precision of the Doppler measurement for the radio frequency subsystem is within $\pm$0.1 mm/s over 10s to several minutes of integration time \citep{Srinivasan07}.

\subsection{Data analysis and dynamic modeling}
\label{analysis}  
We have analyzed one-and-half years of tracking data collected by the \gls{DSN} during the \gls{MGR} orbital period. These data belong to one year of the prime mission and six months of the first extended mission (see Table \ref{data_sum}). The complete data set that was used for the analysis is available on the Geoscience node\footnote{\url{http://pds-geosciences.wustl.edu/messenger/}} of the NASA\textquoteright s \gls{PDS}. For precise orbit determination, all available observations were analyzed with the help of the \gls{GINS} software, which was developed by the \gls{CNES} in collaboration with \gls{ROB}. \gls{GINS} numerically integrates the equations of motion and the associated variational equations. It simultaneously retrieves the physical parameters of the force model using an iterative least-squares technique. 

\subsubsection{Dynamic modeling and orbit determination processes.}
\label{dynamic}
The precise orbit determination is based on a full dynamical approach. The dynamic modeling includes gravitational (gravitational attraction of Mercury, third-body gravity perturbations from the Sun and other planets, and relativistic corrections) and nongravitational (solar radiation pressure; Mercury radiation pressure) forces that are acting on the spacecraft. These forces have been taken into account in the force budget of \gls{MGR}. The latest spherical harmonic model \citep{Smith12} of Mercury\textquoteright s gravity field, HgM002\footnote{\url{http://pds-geosciences.wustl.edu/missions/messenger/rs.htm}} developed up to degree and order 20, and the associated Mercury\textquoteright s orientation model \citep{Margot09} have been considered for precise computation. 

The measurement (Doppler and range) models and the light time corrections that are modeled in \gls{GINS} correspond to the formulation given by \cite{Moyer}. During computations, \gls{DSN} station coordinates were corrected from the Earth\textquoteright s polar motion, from solid-Earth tides, and from the ocean loading. In addition to these corrections, radiometric data have also been corrected from tropospheric propagation through the meteorological data\footnote{\url{http://pds-geosciences.wustl.edu/messenger/mess-v_h-rss-1-edr-rawdata-v1/messrs_0xxx/ancillary/wea/}} (pressure, temperature, and humidity) of the stations.

The complex geometry of the \gls{MGR} spacecraft was treated as a combination of flat plates arranged in the shape of a box, with attached solar arrays, the so-called \textit{Box-Wing} macro-model. The approximated characteristics of this macro-model, which includes cross-sectional area and specular and diffuse reflectivity coefficients of the components, were taken from \citep{Vaughan02}. In addition to the  macro-model characteristics, orientations of the spacecraft were also taken into account. The attitude of the spacecraft and of its articulated panels in inertial space were defined in terms of quaternions. The approximate value of these quaternions was extracted from the C-kernel\footnote{\url{ftp://naif.jpl.nasa.gov/pub/naif/pds/data/mess-e_v_h-spice-6-v1.0/messsp_1000/data/ck/}} system of the SPICE \gls{NAIF} software. The macro-model and its orientation allowed calculation of the nongravitational accelerations that are acting on the \gls{MGR} spacecraft due to the radiation pressure from Sun and Mercury (albedo and thermal infrared emission).

For orbit computation and parameters estimation, a multi-arc approach was used to get independent estimates of the \gls{MGR} accelerations. In this method, we integrated the equations of motion using the time-step of 50s then, and orbital fits were obtained from short data arcs fitted over the observations span of one day using an iterative process. The short data arcs of one day were chosen to account for the model imperfections. To initialize the iteration, the initial position and velocity vectors of \gls{MGR} were taken from the SPICE \gls{NAIF} spk-kernels\footnote{\url{ftp://naif.jpl.nasa.gov/pub/naif/pds/data/mess-e_v_h-spice-6-v1.0/messsp_1000/data/spk/}}.

\begin{table*} 
\caption{Summary of the Doppler and range tracking data used for orbit determination.}
\centering
\begin{center}
\renewcommand{\arraystretch}{1.4}
\small
\begin{tabular}{ c c c c c c c}
 \hline
 \hline  
    Mission       &    Begin date      &    End date      &  Number of        & Number of        & Number of   \\
         phase    &     dd-mm-yyyy    &     dd-mm-yyyy    & 2-way Doppler   & 3-way Doppler  & range           \\
 \hline    
  Prime & 17-05-2011  & 18-03-2012  &  2108980 & 184138 & 11540 \\
  Extended & 26-03-2012 & 18-09-2012  & 1142974 & 23211 & 5709 \\  
  \hline    
\end{tabular}
\end{center}
\label{data_sum}
\end{table*}

\subsubsection{Solve-for parameters}
\label{parameter}
An iterative least-squares fit was performed on the complete set of Doppler- and range-tracking data arcs that correspond to the orbital phase of the mission using an INPOP10e \citep{Fienga13} planetary ephemeris\footnote{\url{http://www.imcce.fr/inpop/}}. We have processed data from May  17 2011 to September 18 2012 excluding the periods of the maneuvers. A summary of these tracking data is given in Table \ref{data_sum}. 
MESSENGER fires small thrusters to perform momentum dump maneuver (MDM) for reducing the spacecraft angular momentum to a safe level. Normal operations (during orbital periods) includes only one commanded momentum dump every two weeks. In addition, orbit correction maneuvers (OCM) were also performed (typically once every Mercury year, 88 Earth days) to maintain the minimum altitude below 500 kilometers. Such large intervals between the MESSENGER maneuvers facilitate the orbit determination. The data arcs that correspond to the maneuver epochs are thus not included in the analysis. The total 440 one-day data arcs were then used for the analysis.

Several parameters were estimated during orbit computation: spacecraft state vectors at the start epoch of each data arc, for a total of 440$\times$6=2640 parameters; one scale factor per data arc for taking into account the mismodeling of the solar radiation force (total of 440 parameters); one Doppler bias per arc for each \gls{DSN} station to account for the systematic errors generated by the devices at each tracking station (total of $\sum_1^{440}1\times$n parameters, where $n$ is the number of stations participating in the data arc); one station bias per arc for each DSN station to account for the uncertainties on the \gls{DSN} antenna center position or the instrumental delays (total of $\sum_1^{440}1\times$n parameters); and one range bias per arc for ranging measurements to account for the systematic geometric positions error (ephemerides bias) between the Earth and the Mercury (total of 440 parameters).
  
\subsection{Orbit determination}
\label{orbit_results}  
 \begin{figure*}[]
\begin{center}\includegraphics[width=15cm]{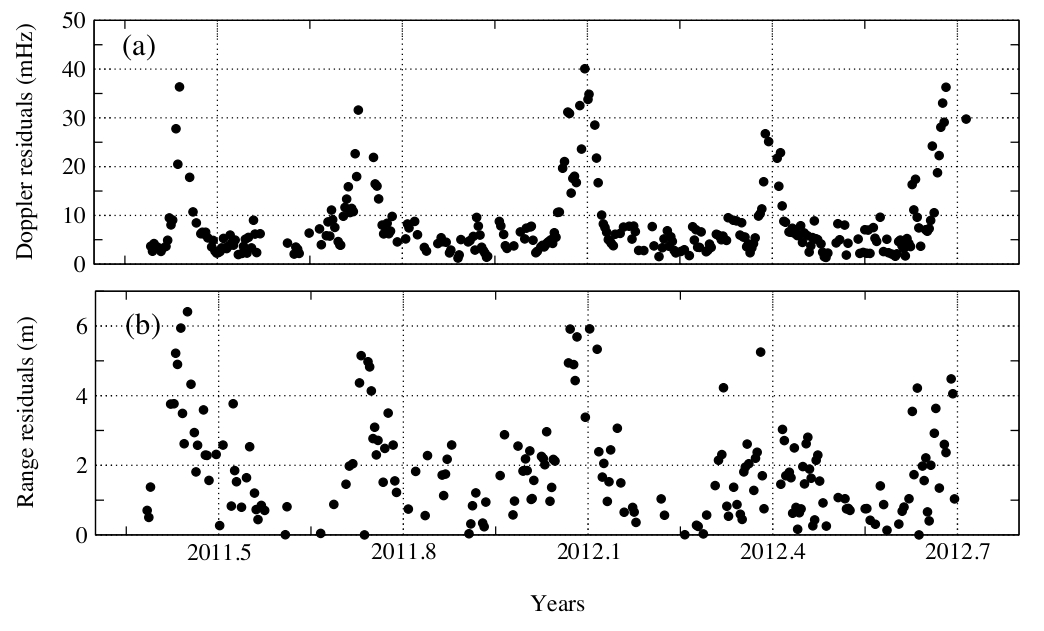}\end{center}
\caption{Quality of the MESSENGER orbit in terms of rms values of the post-fit residuals for each one-day data arc: (a) two- and three-way Doppler given in millihertz (multiply by 0.0178 to obtain residuals in mm/s), and (b) two-way range given in meters. }
\label{resd}
\end{figure*}
\subsubsection{Postfit residuals}
\label{postfit}
The \gls{rms} values of the post-fitted Doppler and range residuals give some indication about the quality of the orbit fit and the estimated parameters. Moreover, the quality of the used parameters associated to the physical model can also be judged from these residuals. Figure \ref{resd} illustrates the time history of the residuals estimated for each measurement type. In this figure, panel $a$ represents the \gls{rms} values of the two- and three-way Doppler residuals that were obtained for each data arc and are expressed in millihertz (mHz). As Mercury has shorter orbit than other planets, it experiences five superior conjunctions (when the Earth, the Sun and the spacecraft lie on the same line, with the spacecraft located on the opposite side of the Sun with respect to Earth) during the time interval covered by the analysis. Because of a lack of modelisation of the solar corona perturbations within the GINS software, no model of solar plasma was applied during the computations of the \gls{MGR} orbit. The peaks shown in Fig. \ref{resd}, therefore, demonstrate the clear effect of the solar conjunctions on the typical fit to the Doppler and range data residuals.

Excluding the solar conjunction periods (about 100 data arcs), when Sun-Earth-Probe angle remained below 10$^\circ$, an average value of Doppler residuals has been found to be approximately 4.8$\pm$2.2 mHz ($\sim$0.09$\pm$0.04 mm/s), which is comparable with values given by \citep{Smith12,Stanbridge11,Srinivasan07}. The mean value of the estimated Doppler bias for each \gls{DSN} station tracking pass was found to be very small (a few tenths of mHz), which is lower than the Doppler post-fit residuals for each data arc. It demonstrated that we have no large bias in the modeling of the Doppler shift measurements at each tracking station.

The range measurements were also used to assist in fitting the Doppler data for a precise orbit determination. Panel $b$ of Fig. \ref{resd} represents the \gls{rms} values of two-way range residuals that were obtained for each data arc. An average value of these range residuals is 1.9$\pm$1.4 m, which is comparable with the values given in \citep{Srinivasan07}.

 \begin{figure*}[]
\begin{center}\includegraphics[width=15cm]{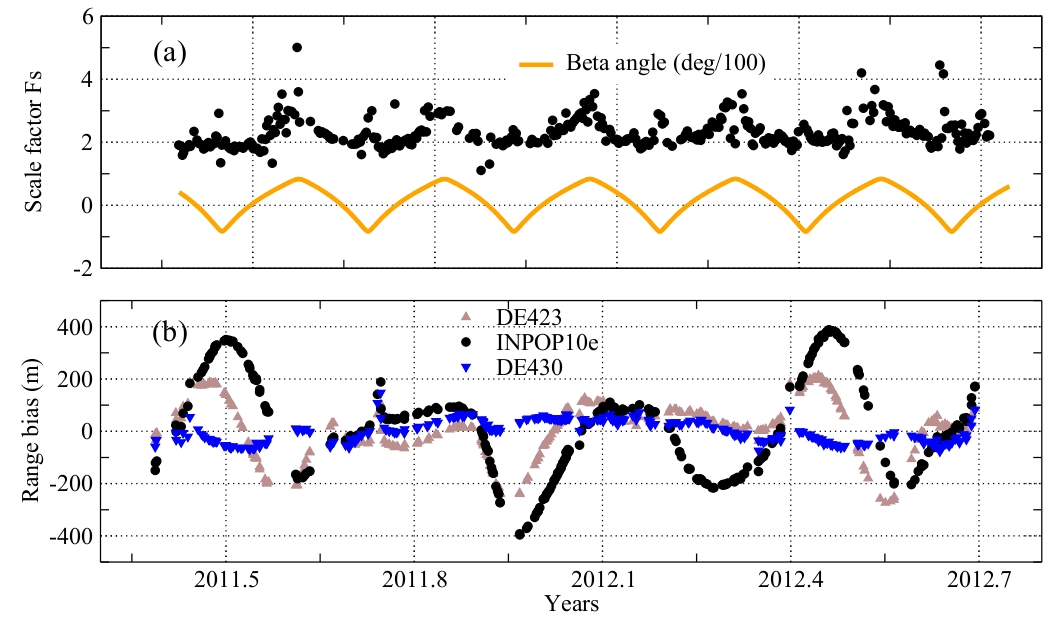}\end{center}
\caption{History of the fitted scale factor and estimated range bias: (a) scale factor (for solar radiation acceleration) fitted over each one-day data arc to account inaccuracy in the force model, and (b) one-way range bias, represent the systematic error in the Earth-Mercury positions, estimated for each one-day arc using INPOP10e ($\bullet$), DE423 ($\blacktriangle$), and DE430 ($\blacktriangledown$) planetary ephemerides.}
\label{bias_fs}
\end{figure*}
\subsubsection{Scale factor and range bias}
\label{scale}
We fitted one scale factor per data arc for the solar radiation force to account the inaccuracy in the force model. Panel $a$ of Fig. \ref{bias_fs} represents the time history of these scale factors. These scale factors are overplotted with the \textit{beta angle}, which is the angle between \gls{MGR} orbital plane and the vector from the Sun direction. The variation in the \gls{MGR} orbital plane (beta angle) relative to the Sun occurs as Mercury moves around the Sun. For example, at 10$^\circ$, 100$^\circ$, and 180$^\circ$ of Mercury's true anomaly, the corresponding beta angles are 83$^\circ$, 0$^\circ$, and -78$^\circ$, respectively \citep{Ercol}.  At 0$^\circ$ beta angle, the spacecraft travels directly between the Sun and the planet, while at 90$^\circ$, the spacecraft is in sunlight 100\% of the time. As one can see from panel $a$ of Fig. \ref{bias_fs}, the solar pressure coefficients have variations that approximately follow those of the beta angle. This implies that, whenever \gls{MGR} orbital plane approaches the maximum beta angle, it is fully illuminated by direct sunlight (no shadow affect). To protect the spacecraft from the direct sunlight, the automatic orientation of the solar panels therefore balances the need for power and the temperature of the surface of the panel. Thus, imperfection in the modeling of these orientations is then taken care of by the scale factor to reduce the error in the computation of solar radiation pressure (see Fig. \ref{bias_fs}). The fitted scale factor for solar radiation pressure is, therefore, typically in the range of about 2.1$\pm$0.5. This value is nearly twice the \textit{a priori} value and it reflects the imperfection in the force model due to the approximate representation of the macro-model.

Panel $b$ of Fig. \ref{bias_fs} illustrates the one-way range bias estimated for the ranging measurements for each data arc. These biases represent the systematic uncertainties in the Earth-Mercury geometric positions. The black ($\bullet$), brown ($\blacktriangle$) and blue ($\blacktriangledown$) bullets in this figure correspond to INPOP10e \citep{Fienga13}, DE423 \citep{DE423}, and DE430 \citep{DE430}, respectively. An average value of these range bias for INPOP10e, DE423 and DE430 is 21$\pm$187 m, 15$\pm$105 m, and -0.5$\pm$42 m, respectively. This range bias is then used in the planetary ephemerides to fit the dynamical modeling of the planet motions (see Sec. \ref{inpop}). Thus, \gls{MGR} ranging measurements were used to reconstruct the orbit of Mercury around the Sun. The improved planetary ephemeris, INPOP13a (see Sec. \ref{inp13}) was then used to re-analyze the \gls{MGR} radiometric data to study the impact of planetary ephemeris over the computation of \gls{MGR} orbit and associated parameters (see Sec. \ref{mgr_eph}). 

\subsubsection{Spacecraft transponder group delay calibration}
Planetary ephemerides are a good tool for testing the gravity model and \gls{GR} \citep{Fienga2011} and performing solar corona studies \citep{verma12}. Moreover, it is also possible to calibrate the transponder group delay from the planetary ephemeris. The spacecraft receives and transmits the signal to the Earth station through the on-board transponder, which causes the time delay in the range measurements. This delay varies from one spacecraft to another depending on the radio frequency configuration. Usually an average value for this delay is measured at different occasions on the ground before launch. However, the group delay is not perfectly stable and can fluctuate by a few ns, depending upon variations in a number of parameters such as temperature and signal strength. 

For \gls{MGR}, we estimated this group delay with the planetary ephemeris. This procedure becomes an alternate method of testing the procedure and quality of the orbit fit by comparing estimated group delay with the delay tested on the ground. { Since the transponder delay does not affect the Doppler measurements we were therefore, able to compute the precise orbit of the spacecraft without considering the transponder delay in the range measurements. With this configuration, we then reanalyzed the entire radio tracking data (see Table \ref{data_sum}). To check the precision on the knowledge of the spacecraft orbit, we compared the radial, along-track, and cross-track components of the orbit for each data arc with the solution obtained in Section \ref{postfit}. An average rms value of radial, along-track, and cross-track difference is 0.015m, 0.16m, and 0.19m, respectively. Less than a meter level of differences in the orbit implies that the transponder delay has negligible impact on the orbit, since the spacecraft orbit is mostly constrained by the Doppler tracking data. However, there is a dramatic change in the estimation of range bias, which now includes ephemeris bias plus the bias due to the transponder delay. Using these range biases  to fit the planetary ephemeris, we found a clear off-set in the Earth-Mercury geocentric distances of about 410$\pm$20 m (two-way) during the orbital period of the \gls{MGR}.} This estimation of transponder delay is compatible with the one found during ground testing, which ranged from 1,356.89 ns ($\sim$ 407 m) to 1,383.74 ns ($\sim$ 415 m) \citep{Srinivasan07}. Thus these results also suggested that there is not a large error in the spacecraft and the planetary orbit fit procedure. 


\section{Improvement of planetary ephemeris, INPOP}
\label{inpop}

\label{gen}
Since 2003, INPOP planetary ephemerides have been built on a regular basis and provided to users thought the IMCCE website \url{www.imcce.fr/inpop}. The INPOP10e ephemeris was the latest release \citep{Fienga13} that was delivered as the official Gaia mission planetary ephemerides used for the navigation of the satellite as well as for the analysis of the data. Specific developments and analysis were done for the Gaia release such as the TCB time-scale version or an accurate estimation of the INPOP link to ICRF. With the delivery of the MESSENGER radio science data, a new opportunity was offered to improve drastically our knowledge of the orbit of Mercury and to perform tests of gravity at a close distance from the Sun. 

The use of the 1.5 year range measurements deduced from the previous analysis ( see section \ref{messenger}) is then a crucial chance to obtain better knowledge over the $\sim$0.3 year Mercury orbit. The accuracy of a few meter for the MESSENGER range data will give big constraints over short period perturbations on Mercury\textquoteright s orbit. The five flyby positions obtained with Mariner and the MESSENGER flybys will still be significant for the measurements of long period (10 or more years) perturbations (see Figure \ref{diffbg}).  Only the addition of the Bepi-Colombo range data will be able to disentangle such long period effects. 


\subsection{INPOP13a}
\label{inp13}

The constants and dynamical modeling used for constructing the new ephemerides, INPOP13a, are similar to INPOP10e. A complete adjustment of the planet initial conditions (including Pluto and the Moon), the mass of the Sun, the oblateness of the Sun, the ratio between the mass of the Earth and the Moon, and 62 asteroid masses is operated. Values of the obtained parameters are given in Tables \ref{paramfita} and \ref{mass0}. Even if Mercury is not directly affected by the main belt asteroids, the use of the range measurements between MESSENGER and the Earth does have an impact on the Earth's orbit and then could bring some information on asteroid masses perturbing the Earth orbit. On Table \ref{mass0}, we only gave the masses that are significantly different from those obtained with INPOP10e and inducing detectable signatures below five meters. { These masses are also compared with the \cite{Konopliv11} on the same table.} The masses of the biggest objects differ from the two ephemerides inside their two-sigma error bars, and one can notice the new determination of the mass of (51) Nemausa inducing slightly bigger perturbations on Mercury (7 meters) and Venus (8 meters) geocentric distances than on Mars (5 meters). 

 \begin{figure*}[!ht]
\begin{center}\includegraphics[width=12cm]{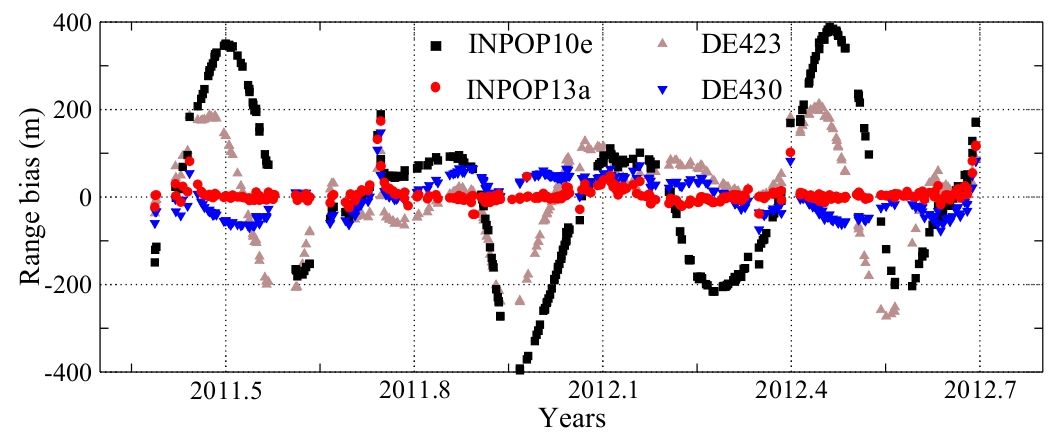}\end{center}
\caption{MESSENGER one-way range residuals obtained with INPOP13a, INPOP10a, DE423, and DE430.}
\label{bias}
\end{figure*}

\begin{table*}[!ht]
\caption{Values of parameters obtained in the fit of INPOP13a and INPOP10e to observations including comparisons to DE423 and DE430.}
\begin{center}
\begin{tabular}{l c c c c}
\hline
&  INPOP13a & INPOP10e & DE423 & DE430 \\
&    $\pm$ 1$\sigma$ & $\pm$ 1$\sigma$ & $\pm$ 1$\sigma$ & $\pm$ 1$\sigma$ \\
\hline
(EMRAT-81.3000)$\times$ 10$^{-4}$ &  (5.770 $\pm$ 0.020) & (5.700 $\pm$ 0.020) & (5.694 $\pm$ 0.015) & (5.691$\pm$0.024)\\

J$_{2}$$^{\odot}$ $\times$ 10$^{-7}$ & (2.40 $\pm$ 0.20) & (1.80 $\pm$ 0.25)  & 1.80 & (2.1$\pm$0.7) \\
\hline
GM$_{\odot}$ - 132712440000 [km$^{3}.$ s$^{-2}$]&  (48.063 $\pm$ 0.4) & (50.16 $\pm$ 1.3) & 40.944 & 41.94 \\
AU - 1.49597870700 $\times$ 10$^{11}$ [m] & 9.0 & 9.0 & (-0.3738 $\pm$ 3 ) & 0\\
\hline
 \end{tabular}
\end{center}
\label{paramfita}
\end{table*}

\begin{table*}[!ht]
\caption{Asteroid masses obtained with INPOP13a, significantly different from values found in INPOP10e, and inducing a change in the Earth-planets distances smaller than 5 meters over the fitting interval. The uncertainties are given at 1 published sigma and compared with \cite{Konopliv11}.}
\begin{center}
\begin{tabular}{c c c c c c}
\hline
IAU designation	& INPOP13a	&  INPOP10e	& \cite{Konopliv11} \\
number & $10^{12}$ x M$_{\odot}$	&  $10^{12}$ x M$_{\odot}$ &  $10^{12}$ x M$_{\odot}$ \\
\hline
    1 &  468.430 $\pm$    1.184     &  467.267 $\pm$    1.855  & 467.90 $\pm$ 3.25 \\
    2 &  103.843 $\pm$    0.982     &  102.654 $\pm$    1.600   & 103.44 $\pm$ 2.55 \\
    9 &    3.637 $\pm$    0.400     &    4.202 $\pm$    0.670    & 3.28 $\pm$ 1.08 \\
   15 &   14.163 $\pm$    0.555     &   15.839 $\pm$    0.950    & 14.18 $\pm$1.49 \\
   16 &   11.212 $\pm$    1.373     &   12.613 $\pm$    2.208    & 12.41 $\pm$ 3.44 \\
   19 &    5.182 $\pm$    0.342     &    4.892 $\pm$    0.513    & 3.20 $\pm$ 0.53 \\
   46 &    3.076 $\pm$    0.446     &    3.525 $\pm$    0.743    & -- \\
   51 &    3.287 $\pm$    0.485     &    0.009 $\pm$    0.004    & -- \\
   65 &    8.789 $\pm$    2.266     &    4.210 $\pm$    0.863    & -- \\
   78 &    1.486 $\pm$    0.504     &    2.562 $\pm$    0.574    & -- \\
  105 &    2.070 $\pm$    0.365     &    3.046 $\pm$    0.635    & -- \\
  106 &    3.369 $\pm$    0.408     &    3.870 $\pm$    0.411    & -- \\
  134 &    3.451 $\pm$    0.595     &    1.014 $\pm$    0.368    & -- \\
  194 &    4.872 $\pm$    0.452     &    5.601 $\pm$    0.636    & -- \\
   324 &    5.087 $\pm$    0.189     &    4.769 $\pm$    0.435    & 5.34 $\pm$ 0.99 \\
\hline
  \end{tabular}
\end{center}
\label{mass0}
\end{table*}

Table \ref{omctab1} gives the postfit residuals obtained with INPOP13a and compared with those obtained with INPOP10e. One can see a noticeable improvement in Mercury's orbit over all the periods of the fit including direct radar observations. The result is of course more striking for MESSENGER range measurements that were deduced from section \ref{messenger}, and not used for the fit of INPOP10e. In this particular case, the improvement reaches a factor of almost 16 on the estimation of the distance between Mercury and the Earth (see Figure \ref{bias}). 
The extrapolated residuals given in Table \ref{omctab1} are not really significant since INPOP10e was fitted over a very similar interval of time ending at about 2010.4 when INPOP13a was fitted up to 2011.4.

\begin{figure*}[]
\begin{center}\includegraphics[width=9.5cm]{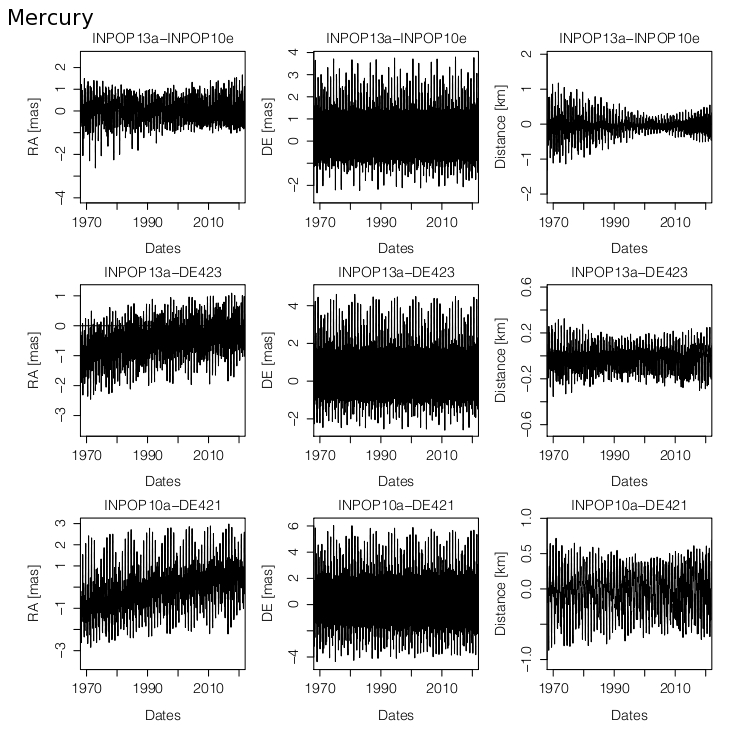}\includegraphics[width=9.5cm]{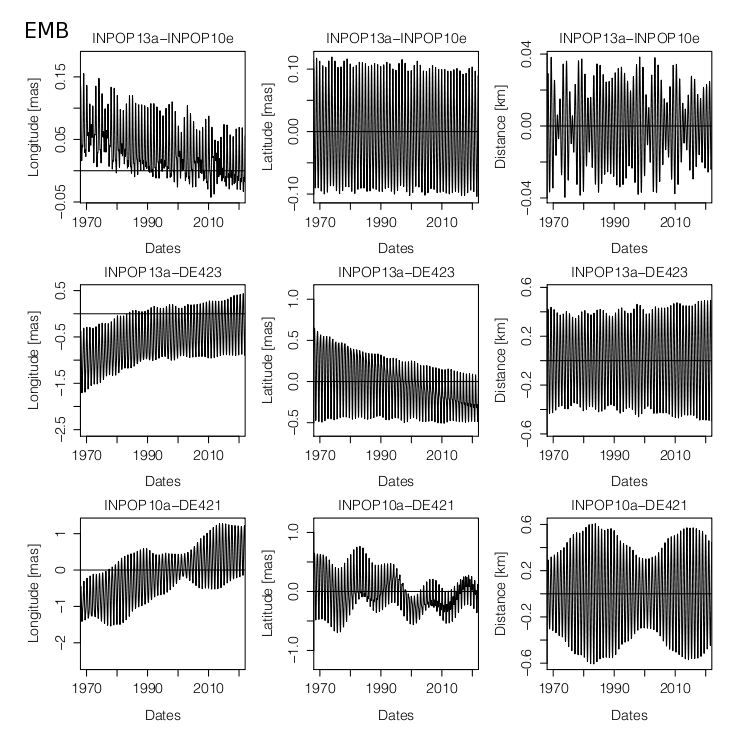}\end{center}
\caption{In Mercury panel, differences in geocentric Mercury right ascension (RA), declination (DE) and distances between INPOP13a, INPOP10e and DE423. In EMB panel, differences in BCRF longitudes, latitudes and distances of the EMB  between INPOP13a, INPOP10e and DE423. Differences between INPOP10a and DE421 are also given.}
\label{diff}
\end{figure*}

Figure \ref{diff} plots the differences between INPOP13a, INPOP10e and DE423 for Mercury geocentric right ascension, declination, and distance, and the Earth-Moon barycenter longitudes, latitudes, and distances in the BCRS. These differences give estimations of the internal accuracy of INPOP13a. By comparison, the same differences between INPOP10a and DE421 are also plotted. They present the improvements reached since INPOP10a, clearly noticeable for the Mercury geocentric distances (a factor two between INPOP13a-DE423 and INPOP10a-DE421). They are less impressive for the EMB; however, one can notice that the clear systematic trend in the INPOP10a-DE423 barycentric distances of the EMB is removed in INPOP13a-DE423. The fact that the differences between INPOP13a and INPOP10e are smaller than the differences to DE ephemerides is mainly discussed in \cite{Fienga2011} and \cite{Fienga13} by a different method of computing the orbit of the Sun relative to the solar system barycenter, as well as a different distribution of planetary and asteroid masses.

In conclusion, INPOP13a shows an important improvement in the Mercury orbit especially during the MESSENGER orbital and flyby phases of the mission. The improvement over the EMB orbit in the BCRS is less important but still a systematic trend noticeable in the  EMB barycentric distance differences between INPOP10a and DE421 seems to be removed in the new comparisons.

\begin{table*}[!ht]
\caption{Statistics of the residuals obtained after the INPOP13a fit. For comparison, means and standard deviations of residuals obtained with INPOP10e are given. The label {\it{GINS range}} indicates that the corresponding data set was obtained after orbit reconstruction of the spacecraft in using the GINS software. For MGS, see \cite{verma12}. }
\begin{center}
\begin{threeparttable}
\begin{tabular}{l c l c | c c | c c} 
\hline
Type of data & & Nbr & Time Interval & \multicolumn{2}{c}{INPOP13a}& \multicolumn{2}{c}{INPOP10e}  \\
& & & & mean & 1-$\sigma$ & mean & 1-$\sigma$ \\
\hline
Mercury & range [m]& 462 &  1971.29 - 1997.60 &    -108 &    866 & -45 & 872  \\
{Mercury Messenger} & {GINS range [m]}& {314} &  {2011.39 - 2012.69} &       {2.8} &    {12.0} & {15.4} &    {191.8}  \\
{Out from SC\tnote{*}} & {GINS range [m]}& {267} &  {2011.39 - 2012.66} &      {-0.4} &      {8.4} & {6.2} &    {205.2}  \\
Mercury Mariner & range [m]& 2 &  1974.24 - 1976.21 &    -124 &     56 & -52.5 & 113  \\
Mercury flybys Mess & ra [mas]& 3 &  2008.03 - 2009.74 &       0.85 &      1.35 & 0.73 & 1.48  \\
Mercury flybys Mess & de [mas]& 3 &  2008.03 - 2009.74 &       2.4 &       2.4 & 2.4 & 2.5  \\
Mercury flybys Mess & range [m]& 3 &  2008.03 - 2009.74 &      -1.9 &       7.7 & -5.05 & 5.8  \\
\hline
Venus & VLBI [mas]& 46 &  1990.70 - 2010.86 &       1.6 &      2.6 &1.6 & 2.6  \\
Venus & range [m]& 489 &  1965.96 - 1990.07 &     502 &   2236 &500 & 2235  \\
Venus Vex & range [m]& 24970 &  2006.32 - 2011.45 &       1.3 &     11.9 & 1.1 &     11.9  \\
\hline
Mars & VLBI [mas]& 96 &  1989.13 - 2007.97 &      -0.02 &      0.41 &-0.00 & 0.41  \\
Mars Mex & range [m]& 21482 &  2005.17 - 2011.45 &      -2.1 &     20.6 &-1.3 &     21.5 \\
Mars MGS & GINS range [m]& 13091 &  1999.31 - 2006.83 &      -0.6 &      3.3 &-0.3 & 3.9  \\
Mars Ody & range [m]& 5664 &  2006.95 - 2010.00 &       1.6 &      2.3 &0.3 & 4.1  \\
Mars Path & range [m]& 90 &  1997.51 - 1997.73 &       6.1 &     14.1 &-6.3 & 13.7  \\
Mars Vkg & range [m]& 1257 &  1976.55 - 1982.87 &      -0.4 &     36.1 &-1.4 & 39.7  \\
\hline
Jupiter& VLBI [mas]& 24 &  1996.54 - 1997.94 &      -0.5 &     11.0 &-0.3 & 11.0  \\
Jupiter Optical& ra [mas]& 6532 &  1914.54 - 2008.49 &      -40 &      297 &-39 & 297  \\
Jupiter Optical& de [mas]& 6394 &  1914.54 - 2008.49 &      -48 &      301 &-48 & 301  \\
Jupiter flybys & ra [mas]& 5 &  1974.92 - 2001.00 &       2.6 &      2.9 &2.4 & 3.2 \\
Jupiter flybys & de [mas]& 5 &  1974.92 - 2001.00 &     -11.0 &      11.5 &-10.8 & 11.5  \\
Jupiter flybys & range [m]& 5 &  1974.92 - 2001.00 &   -1065 &    1862 &-907 & 1646  \\
\hline
Saturne Optical& ra [mas]& 7971 &  1913.87 - 2008.34 &      -6 &      293 &-6 & 293  \\
Saturne Optical& de [mas]& 7945 &  1913.87 - 2008.34 &      -12 &      266 &-2 & 266  \\
Saturne VLBI Cass & ra [mas]& 10 &  2004.69 - 2009.31 &       0.19 &      0.63 &0.21 & 0.64  \\
Saturne VLBI Cass & de [mas]& 10 &  2004.69 - 2009.31 &       0.27 &      0.34 &0.28 & 0.33  \\
Saturne Cassini & ra [mas]& 31 &  2004.50 - 2007.00 &       0.8 &      3.4 &0.8 & 3.9 \\
Saturne Cassini & de [mas]& 31 &  2004.50 - 2007.00 &       6.5 &       7.2 &6.5 & 7.2  \\
Saturne Cassini & range [m]& 31 &  2004.50 - 2007.00 &      -0.010 &      18.44 &-0.013 & 18.84  \\
\hline
Uranus Optical & ra [mas]& 13016 &  1914.52 - 2011.74 &       7 &      205 & 7 & 205 \\
Uranus Optical& de [mas]& 13008 &  1914.52 - 2011.74 &      -6 &      234 &-6 & 234  \\
Uranus flybys & ra [mas]& 1 &  1986.07 - 1986.07 &      -21 &      &-21 &  \\
Uranus flybys & de [mas]& 1 &  1986.07 - 1986.07 &      -28 &       &-28 &  \\
Uranus flybys & range [m]& 1 &  1986.07 - 1986.07 &      20.7 &      & 19.7 &  \\
\hline
Neptune Optical& ra [mas]& 5395 &  1913.99 - 2007.88 &       2 &      258 &0.0 & 258  \\
Neptune Optical& de [mas]& 5375 &  1913.99 - 2007.88 &      -1 &      299 &-0.0 & 299  \\
Neptune flybys & ra [mas]& 1 &  1989.65 - 1989.65 &      -12 &      &-12 &  \\
Neptune flybys & de [mas]& 1 &  1989.65 - 1989.65 &      -5 &       &-5 &  \\
Neptune flybys & range [m]& 1 &  1989.65 - 1989.65 &      66.8 &       &69.6 &  \\
\hline
Pluto Optical& ra [mas]& 2438 &  1914.06 - 2008.49 &      -186 &      664 &34 & 654  \\
Pluto Optical& de [mas]& 2461 &  1914.06 - 2008.49 &      11 &      536 &7 & 539  \\
Pluto Occ & ra [mas]& 13 &  2005.44 - 2009.64 &       6 &      49 &3 & 47  \\
Pluto Occ & de [mas]& 13 &  2005.44 - 2009.64 &      -7 &      18 &-6 & 18  \\
Pluto HST & ra [mas]& 5 &  1998.19 - 1998.20 &      -42 &      43 &33 & 43  \\
Pluto HST & de [mas]& 5 &  1998.19 - 1998.20 &       31 &      48 &28 & 48  \\
\hline
\hline
Venus Vex\tnote{**} & range [m]& 2827 &  2011.45 - 2013.00 &      51 &    124 & 52 &    125 \\
Mars Mex\tnote{**} & range [m]& 4628 &  2011.45 - 2013.00 &      -3.0 &     11.5 & 4.2 &     27.5  \\
\hline
\end{tabular}
\begin{tablenotes}
\item[*] Solar corona period
\item[**] Extrapolation period 
\end{tablenotes}
\end{threeparttable}
\end{center}
\label{omctab1}
\end{table*}


\subsection{Reconstruction of MESSENGER orbit with INPOP13a}
\label{mgr_eph}
As given in Table \ref{omctab1}, geometric distances between Earth and Mercury are $\sim$16 times better in INPOP13a than the INPOP10e. To analyze the impact of the improvement of the planetary ephemeris on the spacecraft orbit, we reanalyzed the entire one and half years of radioscience data (see Table \ref{data_sum}) using INPOP13a ephemeris. 
 \begin{figure*}[]
\begin{center}\includegraphics[width=15cm]{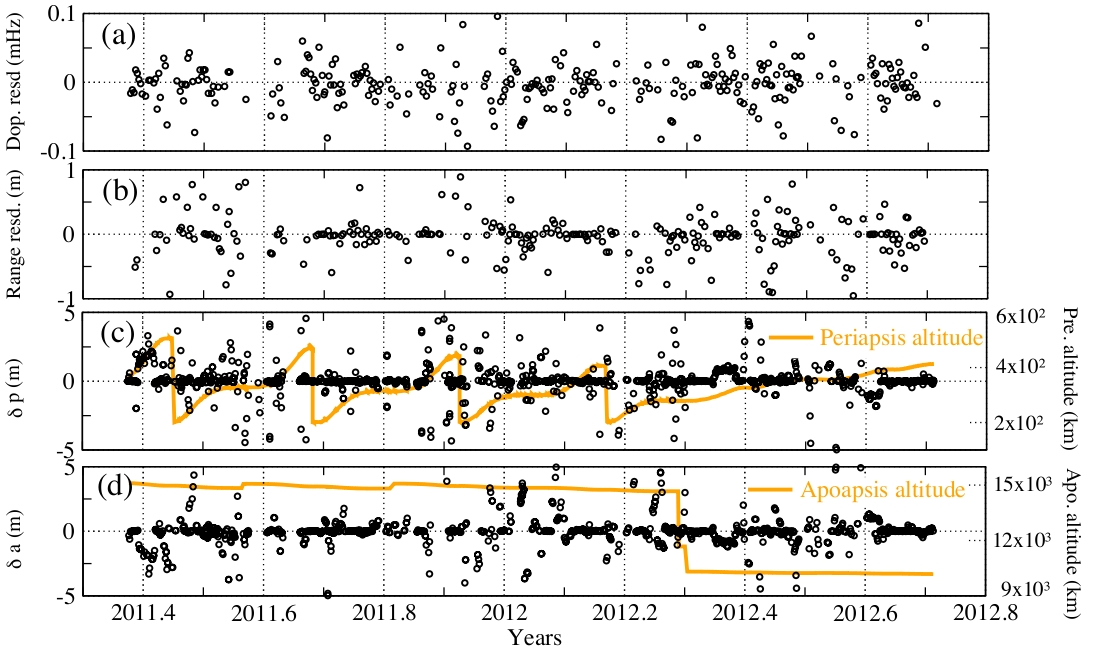}\end{center}
\caption{Comparison between INPOP13a and INPOP10e estimations of MESSENGER orbit: (a) differences in the postfit Doppler residuals; (b) differences in the postfit range residuals; (c) differences in the periapsis altitude $\delta$p; (d) differences in the apoapsis altitude $\delta$a.}
\label{eph_diff_resd}
\end{figure*}
The dynamical modeling and orbit determination process for this analysis are the same as discussed in section \ref{analysis}. To compare the results of this analysis with the one obtained from INPOP10e (see Section \ref{postfit}), the differences in the Doppler and range postfit residuals along with the changes that occurred in the periapsis and apoapsis altitudes of \gls{MGR} are plotted in Fig. \ref{eph_diff_resd}.  

An average value of these differences and its 1$\sigma$ mean dispersion for Doppler, and range postfit residuals was estimated as 0.008$\pm$0.04 mHz and 0.05$\pm$0.3 m, respectively. These values are far below the estimated accuracy of 4.8$\pm$2.2 mHz and 1.9$\pm$1.4 m (see Section \ref{postfit}) for Doppler and range postfit residuals, respectively. In addition to these residuals, we also compared the orbit of \gls{MGR} computed with INPOP13a and INPOP10e ephemerides. The differences in the periapsis $\delta$p and apoapsis $\delta$a altitudes of \gls{MGR} due to the change in planetary ephemeris are plotted in panels $c$ and $d$ of Fig. \ref{eph_diff_resd}. An average and 1$\sigma$ dispersion of $\delta$p and $\delta$a was found as 0.05$\pm$1.2 m and 0.03$\pm$1.2 m, respectively. These values are also far below the required accuracy of 10 m \citep{Srinivasan07} for the \gls{MGR} orbit. This analysis is therefore consistent with the fact that change in the planetary ephemeris during the construction of the spacecraft orbit does not alter the radioscience analysis significantly. 


\section{Test of general relativity}
\label{test}

\subsection{General presentation}
\label{pre}
INPOP13a was built in the framework of \gls{GR} in using the Parametrized Post-Newtonian formalism (PPN). A detailed description of the modeling used for the INPOP ephemerides is given in (\cite{Fienga2011}). Specific relativistic timescales have been set up and integrated in INPOP (TCB, TDB) and mainly two parameters characterized the relativistic formalism in modern planetary ephemerides: the parameter $\beta$ that measures the nonlinearity of gravity and $\gamma$, measuring the deflexion of light. In GR, both are supposed to be equal to 1 and were fixed to 1 for the INPOP13a construction. The GINS software used for the analysis of the radio science data and the reconstruction of the MESSENGER orbit is also coded in the PPN framework, including both $\beta$ and $\gamma$ PPN parameters.

Up to now, general relativity theory (GRT) has successfully described all available observations, and no clear observational evidence against \gls{GR} has been identified. However, the discovery of Dark Energy which challenges GRT as a complete model for the macroscopic universe, and the continuing failure to merge GRT and quantum physics indicate that new physical ideas should be sought. To streamline this search it is indispensable to test GRT in all accessible regimes and to the highest possible accuracy.

Among all possibilities for testing GRT, the tests of the motion and light propagation in the solar system were historically the first ones, and they are still very important since they give highest accuracy since the dynamics of the solar system is well understood and supported by a long history of observational data. 
Concerning the Einstein field equations, the most important framework used for the tests in the solar system is the PPN formalism (such as \cite{Will93}). The PPN formalism is a phenomenological scheme with ten dimensionless parameters covering certain class of metric theories of gravity, among them the $\beta$ and $\gamma$ parameters parts of the INPOP and GINS modelings. The tracking data of space missions give a good possibility to test GRT since the data is very sensitive to the GRT-effects in both dynamics of the spacecraft and signal propagation. However, some factors, such as navigation unknowns (AMDs, solar panel calibrations), planet unknowns (potential, rotation, etc.), effect of the solar plasma, or the correlation with planetary ephemerides limit this sort of gravity test. Dynamics of the solar system are, however, less affected by poorly modeled accelerations and technical unknowns. Up to now, the best constraints for $\beta$ come from the planetary data in INPOP (\cite{Fienga2011}). Constraints on other PPN parameters can be found in \cite{Will06}.  A number of theoretical models predict deviations of PPN parameters that are smaller than current constraints. Typical examples here are certain types of tensor-scalar theories where cosmological evolution exhibits an attractor mechanism towards GRT (\cite{DamourNordtvedt93}) or string-inspired scalar-tensor theories where the scalar field can decouple from matter (\cite{DamourPolyakov94}).

Another phenomenological test concerns the constancy of the Newtonian gravitational constant G in time. A variable G is produced say by alternative theories of gravity such tensor-scalar theory (see e.g. \cite{Damour90} and \cite{2003AnHP....4..347U}) or some models of dark energy (\cite{2009PhRvD..79j4026S}; \cite{2010AIPC.1241..690A}). The ratio is now constrained at the level of 10$^{-13}$ with LLR analysis (\cite{Williams2004}).


\subsection{Estimation of PPN parameters, $\gamma$ and $\beta$}
\label{ppn}
In this section, we propose to use the improvement of Mercury\textquoteright s orbit as an efficient tool for testing the consistency between planetary ephemerides built with MESSENGER radio science data and non-unity PPN parameters.
\subsubsection{Method}
\label{ppn_method}

\begin{figure*}[!ht]
\begin{center}\includegraphics[width=15cm]{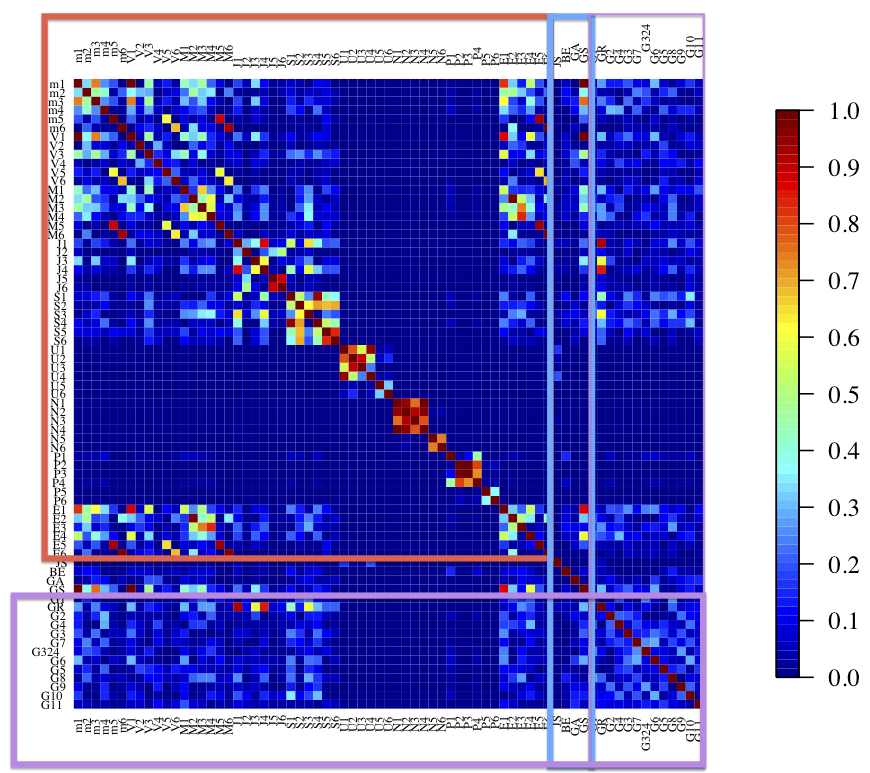}\end{center}
\caption{Correlation between the first 71 (over 343) parameters estimated during the fit of the planetary ephemerides. The red frame frames the correlations related to the initial conditions of planet orbits and the blue rectangle frames the correlations related to the Sun J$_{2}$ ($JS$), the PPN parameters $\beta$ ($BE$) and $\gamma$ ($GA$) and the gravitational mass of the Sun ($GS$). The magenta rectangle frames the correlations related to the gravitational masses of the first most perturbing asteroids including the gravitational mass of the asteroid ring ($GR$). $m_{1},...m_{6}$ expresses the initial conditions of the Mercury orbit in equinoctial coordinates: semi-major axis, mean motion, $k=$ ,$h=$ , $q=$ and $p=$ respectively. The other planet initial conditions are indicated by the first letter of the planet ($V$ for Venus, $M$ for Mars etc...) and by the figures of the corresponding initial conditions as given for Mercury. }
\label{correlations}
\end{figure*}

A first estimation of PPN $\beta$ and $\gamma$ is possible by least square methods during the adjustment of the INPOP planetary ephemerides, and the results are given in Table \ref{paramfitc}. Figure \ref{correlations} gives the correlations between the first 71 over the 343 parameters estimated in the adjustments. As one can see in Fig. \ref{correlations} no correlation greater than 0.3 affects the determination of the PPN parameters $\beta$ and $\gamma$, as well as the fit of the Sun oblateness, when the gravitational mass of the Sun is highly related to the Mercury and to the Earth orbits.

However to go further in the analysis of the uncertainties and the construction of acceptable intervals of violation of GR throught the PPN $\beta$ and $\gamma$, we also  considered the same method as the one that was used and described in \cite{Fienga2011} for determining acceptable intervals of violation of \gls{GR} when the PPN formalism. Small variations in these two parameters near unity  are imposed when constructing alternative planetary ephemerides that are fit over the whole data sets presented in Table \ref{omctab1} and with the same parameters and hypothesis as INPOP13a. A minimum of three iterations in the adjustment process is required for building new ephemerides, and comparisons between these ephemerides and INPOP13a are done to scale up what variations to \gls{GR} are acceptable at the level of uncertainty of the present planetary ephemerides. 

\begin{figure*}[!ht]
\begin{center}\includegraphics[width=8cm]{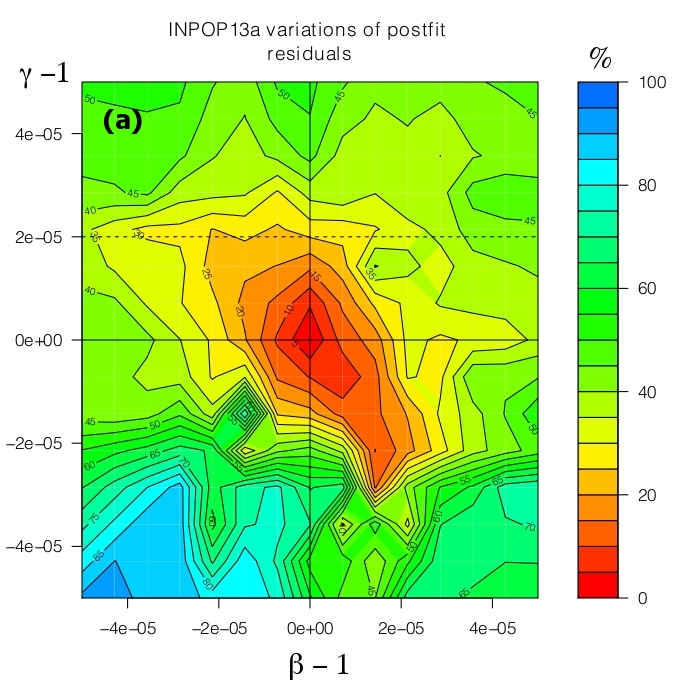}\includegraphics[width=8cm]{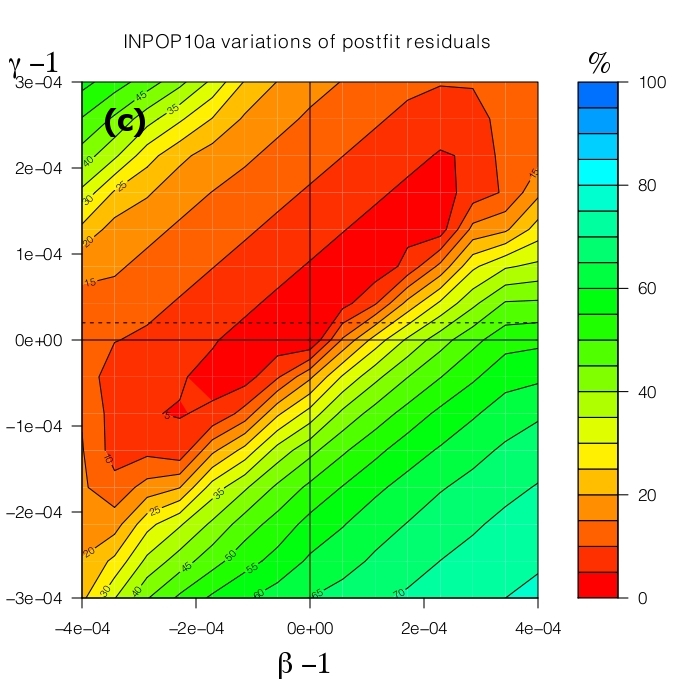}
\includegraphics[width=8cm]{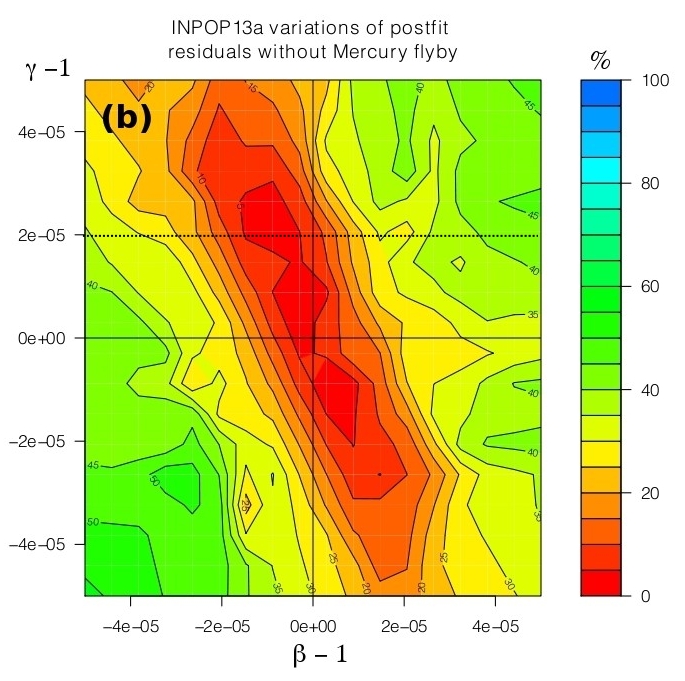}\includegraphics[width=8cm]{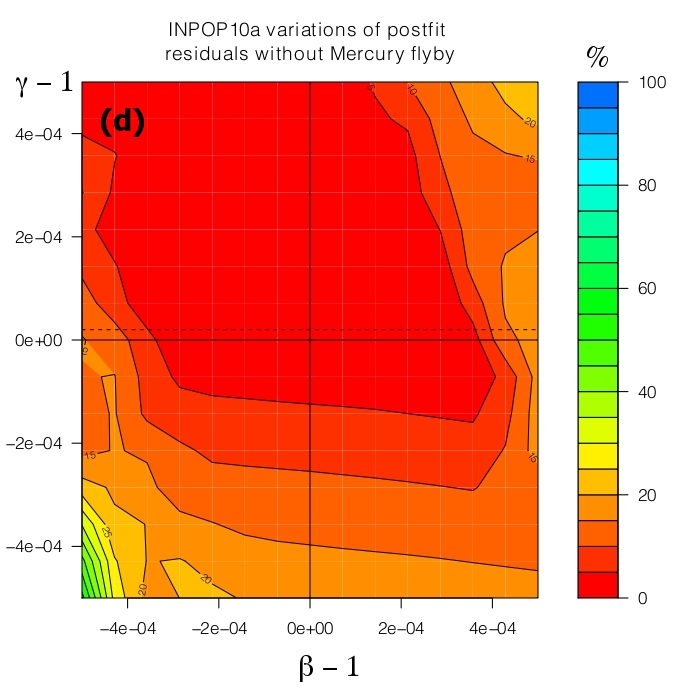}\end{center}
\caption{Variations in postfit residuals obtained for different values of PPN $\beta$ (x-axis) and $\gamma$ (y-axis). Panels (a) and (c) are obtained by considering the variations in the whole data sets when for Panels (b) and (d), variations in the Mercury flyby data (from Mariner and MESSENGER missions) are excluded from the analysis. The dashed line indicates the limit in $\gamma$ given by \cite{Bertotti03}.}
\label{mapfull}
\end{figure*}

\begin{table*}
\caption{Intervals of violation for PPN parameters $\beta$ and $\gamma$ deduced from {{Figure \ref{mapfull} panel (a) labelled INPOP13a and Figure \ref{mapfull} panel (b) labelled INPOP13aWF}}. Values from INPOP10a are extracted from \cite{Fienga2010} and \cite{Fienga2011} with a threshold for the variations of the postfit residuals of 5$\%$ given in Column 4. K11 stands for \cite{Konopliv11}, M08 for  \cite{Muller08}, W09 for \cite{Williams09}, B03 for \cite{Bertotti03}, P13 for \cite{Pitjeva13} and L11 for \cite{Lambert2011}. The {\it{Least squares}} section gives the fitted values of $\beta$ and $\gamma$ at 1$\sigma$ as obtained by a global fit of INPOP presented in section \ref{ppn_method}.}
\begin{center}
\begin{tabular}{l l || c c l}
\hline
Ref.  & $(\beta-1) \times (\gamma-1)$ & INPOP13a & {Limit [$\%$]} & $(\beta-1) \times (\gamma-1)$ \\
 & {$\times$ 10$^{5}$}& & & {$\times$ 10$^{5}$}\\
\hline
& & {All data} & {25} &{($\beta$-1) = (0.2 $\pm$ 2.5)} \\
{INPOP10a} &  {($\beta$-1) = (-6.2 $\pm$ 8.1)} & & & {($\gamma$-1) = (-0.3 $\pm$ 2.5)} \\
&  {($\gamma$-1) = (4.5 $\pm$ 7.5)} & & 10 & {($\beta$-1) = (-0.15 $\pm$ 0.70)} \\
& &  & & {($\gamma$-1) = (0.0 $\pm$ 1.1)} \\
 {K11} &  {($\beta$-1) = (4$\pm$ 24) } & & 5 & {($\beta$-1) = (0.02 $\pm$ 0.12)} \\ 
  &  {($\gamma$-1) = (18 $\pm$ 26) } & & &  {($\gamma$-1) = (0.0 $\pm$ 0.18)} \\
 & & & & \\
{M08-LLR-SEP}* &{($\beta$-1) = (15 $\pm$ 18)} &  Least squares & & $(\beta-1)^{**}$= (1.34 $\pm$ 0.043)  \\
  {W09-LLR-SEP}*  &{($\beta$-1) = (12 $\pm$ 11)} & & & $(\gamma-1)^{**}$=(4.53 $\pm$ 0.540)\\
& & & & \\
  {{B03-CASS}} & {($\gamma$-1) = (2.1 $\pm$ 2.3)} & {No flyby} & 25 & {($\beta$-1) = (-0.5 $\pm$ 4.5)} \\
  {L11-VLB} &  {($\gamma$-1) = (-8 $\pm$ 12 )} & & & {($\gamma$-1) = (12.5 $\pm$ 17.5)} \\
  & &  &  10 & {($\beta$-1) = (0.0 $\pm$ 2.0)} \\
  {P13}& {($\beta$-1) = (-2$\pm$ 3) } & &  & {($\gamma$-1) = (0.5 $\pm$ 3.5)}\\
  & {($\gamma$-1) = (4 $\pm$ 6) } &  & {5} & {($\beta$-1) = (-0.25 $\pm$ 1.25)} \\
  & & & & {($\gamma$-1) = (-0.1 $\pm$ 2.6)}\\
\hline
\end{tabular}
\end{center}
* values obtained for $(\gamma-1)_{\textrm{B03-CASS}}$ \\
$^{**}$ least square results given at 1$\sigma$ 
\label{paramfitc}
\end{table*}

The improvement of Mercury\textquoteright s orbit in INPOP13a justifies these new estimations. Indeed, Mercury played a historical role in testing gravity and \gls{GR} in 1912 (Einstein 1912) and it is still the planet the most influenced by the gravitational potential of the Sun. Its orbit can then lead to the most efficient constraints on  $\beta$, hence on $\gamma$ in the PPN formalism. Before the recent input of MESSENGER flyby and orbital radio science data in the INPOP construction, Mars was the most constraining planet for the PPN parameters (\cite{Fienga2010}). The reason was the long range of high accurate observations on Mars. The implementation of the first MESSENGER flyby data reduces the interval of violation of $\beta$ to 50 $\%$. The first estimation of the $\gamma$ interval of violation was made possible thanks to the gain in uncertainty on the Mercury orbit. With INPOP13a, even better improvement is achieved.
\begin{figure}[!ht]
\begin{center}\includegraphics[width=8cm]{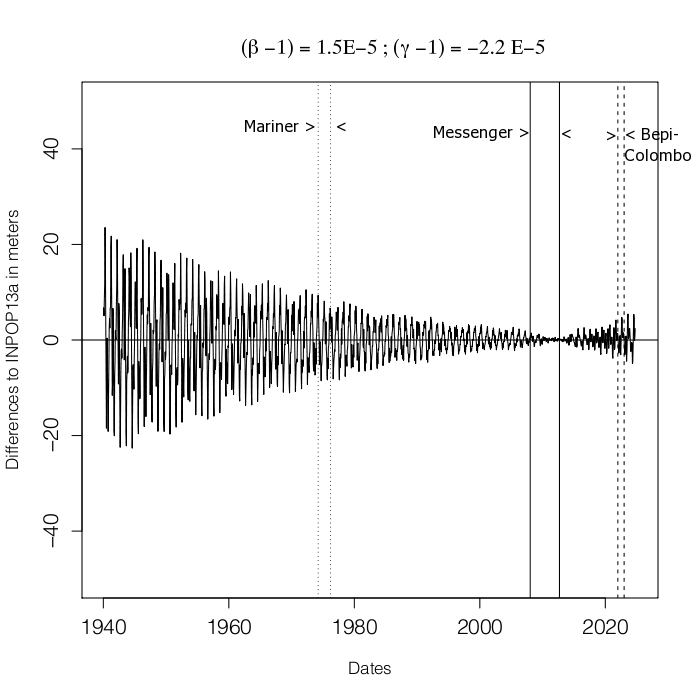}\end{center}
\caption{Differences in geocentric Mercury distances between INPOP13a and a planetary ephemerides built with PPN $\beta$ and $\gamma$ different from 1. The indicated area shows intervals of time corresponding to Mariner observations, MESSENGER and the future Bepi-Colombo.}
\label{diffbg}
\end{figure}
\subsubsection{Results}
\label{ppn_results}

The results obtained by direct least squares fit are presented in Table \ref{paramfitc}. As expected, the estimated uncertainties are very optimistic and a more detailed analysis is done based on the method proposed by \cite{Fienga2011}.

Results obtained in terms of percentages of the variations in postfit residuals between a planetary ephemeris fitted and built up with PPN parameters different from one and INPOP13a are given in Figure \ref{mapfull}. Panel (a) in Figure \ref{mapfull} gives the map of the variations in percent of the full dataset postfit residuals. Panel (b) in Figure \ref{mapfull} gives the same map but without taking the variations of the Mercury flyby data into account. { For Panel (b), the Mercury flyby data are indeed used in the ephemerides fit but not in the analysis of the postfit residuals for testing GR.} The map of Panels (a) and (b) is then dramatically different: where the limits for $\beta$ and $\gamma$ are stringent for the map including the Mariner data, the constraints are greatly enlarged for  these two parameters. These phenomena were expected since the variations in PPN parameters induce long-term perturbations in the geocentric distances of Mercury as one can see in Figure \ref{diffbg}. Panels (c) and (d) are similar to (a) and (b), but they are obtained with ephemeris INPOP10a. In this ephemeris, MESSENGER flyby data were included in its fits but not in the orbital data. By comparing panels (a) and (c), one can see that the use of the MESSENGER orbital data significantly reduce the intervals of violation for both PPN parameters by a factor 10. The same manner, in the most pessimistic case and without considering the Mercury flybys in analysing of the variations in the postfit residual, one can see in Panels (b) and (d) that the improvement of Mercury\textquoteright s orbit is again crucial for reducting the violation intervals of PPN parameters.  

Table \ref{paramfitc} collects the acceptable violation intervals obtained from INPOP10a and INPOP13a. Values extracted from INPOP10a were obtained at 5$\%$ of postfit residual variations \citep{Fienga2011}. {{ With INPOP13a, we extracted values from i) Panel (a) of Figure \ref{mapfull} obtained at 5$\%$, but also at 10$\%$ and 25$\%$ and ii) from Panel (b) of Figure \ref{mapfull} obtained at 5$\%$, which is consistent with the 25$\%$ of intervals extracted from Panel (a)}. }

All given intervals are compatible with GR with an uncertainty at least ten times smaller than our previous results with INPOP10a. In Table \ref{paramfitc}, comparisons to least squares estimations of other planetary ephemerides or Moon ephemerides like \cite{Pitjeva13}, \cite{Konopliv11}, \cite{Muller08}, and \cite{Williams09}, as well as estimations deduced from VLBI observations \cite{Lambert2011}, are also given. The most stringent published constraint for the PPN parameter $\gamma$ has been obtained so far during a dedicated phase of the Cassini mission by \cite{Bertotti03}. This value is compatible with our 25$\%$ estimation { when our 5$\%$ and 10$\%$ estimations give more restrictive intervals of GR violations. 

Confirmations of the results presented in Table \ref{paramfitc} will be obtained by the use of the radioscience data obtained during the future Bepi-Colombo mission. In addition, the recovery of the Mariner flyby data would also be a great help for such confirmations. Unfortunately, the Mariner data seem to have been lost, and access to these data seems to be unrealistic.} 
Indeed as one can see in Figure \ref{diffbg}, perturbations induced by a slight change in the \gls{PPN} parameters (($\beta$-1)= 1.5$\times$10$^{-5}$ and ($\gamma$-1)=-2.2$\times$10$^{-5}$) inducing an effect of about six meters on the Mariner range data (12$\%$) will induce a signature of about the same level at the Bepi-Colombo epoch. With the improved Bepi-Colombo radio science tracking, the expected accuracy in the range measurement is planned to be about 50 centimeters. With such accuracy, detecting the perturbations induced by the same modification of the PPN parameters should be done at 1200$\%$! Two orders of magnitude are expected as a gain in the uncertainty for the $\beta$ and $\gamma$ estimations.

\subsubsection{Impact on MESSENGER orbit}
\label{mgr_ppn} 
 \begin{figure*}[!ht]
\begin{center}\includegraphics[width=15cm]{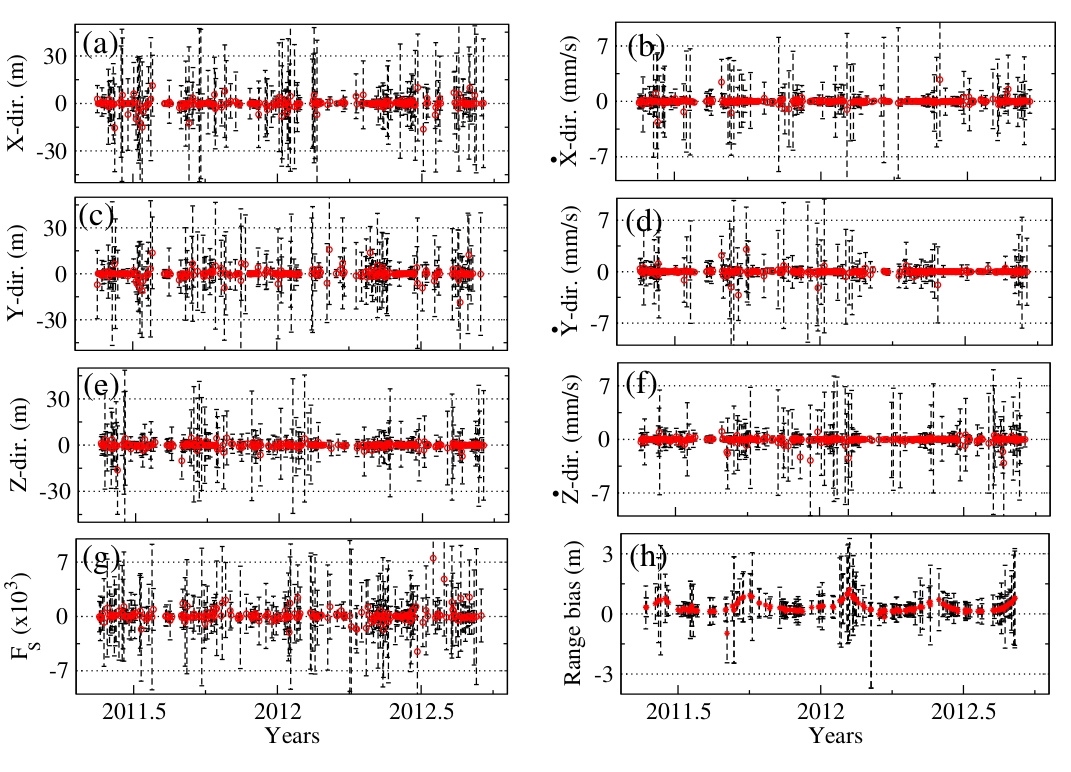}\end{center}
\caption{Differences between solve-for parameters obtained from the solutions SOL$_{Ref}$ and SOL$_{\beta\gamma\neq1}$ (see text):  Panels (a)-(f) represent the changes in the initial state vectors, Panel (g) represents the changes in the scale factors estimated for solar radiation pressure, Panel (h) represents the changes in the estimated range bias. The error bars represent the 1$\sigma$ uncertainties in the estimation of solve-for parameters obtained with the reference solution SOL$_{Ref}$.}
\label{gb_test}
\end{figure*}
As stated previously, the \gls{GINS} software was modeled in the framework of the \gls{PPN} formalism which includes $\beta$ and $\gamma$ parameters. To analyze the combined impact of \gls{PPN} parameters over the \gls{MGR} orbit and the planetary ephemerides construction, we analyzed the entire one and half years of radioscience data again using the \gls{PPN} parameters that are different from unity. The same procedure as described in section \ref{analysis} has been used for reconstructing the \gls{MGR} orbit. Two sets of \gls{MGR} orbits were then built, one with $\beta$ and $\gamma$ equal to unity and the other with $\beta$ and $\gamma$ different from unity (in this case, $\beta$-1 = $\gamma$-1 = 1$\times$10$^{-4}$). Hereafter, the solution obtained from $\beta$ and $\gamma$ equal to unity is referred to as SOL$_{Ref}$ (for the reference solution), and the solution corresponds to $\beta$ and $\gamma$ different from unity referred to as SOL$_{\beta\gamma\neq1}$. 

To maintain consistency in constructing the \gls{MGR} orbit, we used corresponding planetary ephemerides that were built with the same configurations of \gls{PPN} parameters as used for SOL$_{Ref}$ and SOL$_{\beta\gamma\neq1}$. In Fig. \ref{gb_test}, we plotted the differences between solve-for parameters obtained for SOL$_{Ref}$ and SOL$_{\beta\gamma\neq1}$. The error bars shown in the same figure represent the 1$\sigma$ uncertainties in the estimation of solve-for parameters corresponding to SOL$_{Ref}$. From this figure, one can notice the differences in the parameters are always below the 1$\sigma$ uncertainties. The estimated solve-for parameters for SOL$_{\beta\gamma\neq1}$ are analogous to SOL$_{Ref}$, and there is no significant change in the \gls{MGR} orbit due to the change in \gls{PPN} parameters. In contrast, as shown in Fig. \ref{mapfull}, this configuration of \gls{PPN} parameters ($\beta$-1 = $\gamma$-1 = 1$\times$10$^{-4}$) in the construction of planetary ephemerides led to $\sim$65$\%$ of change in the postfit residuals, which shows that, the planetary ephemerides are more sensitive to GR effects. This can be explained from the fitting intervals of the data set. Usually planetary ephemerides are fitted over long intervals of times (see Table \ref{omctab1}) to exhibit long-term effects, while a spacecraft orbit is usually constructed over much shorter intervals (usually one day to a few days) of data arcs to account for the model's imperfections. The short fitting interval of the spacecraft orbit would absorb such effects.

Moreover, it is worth noticing that, unlike state vectors and scale factor F$_S$ (see Panels (a)-(g) of Fig. \ref{gb_test}), the range bias differences between SOL$_{Ref}$ and SOL$_{\beta\gamma\neq1}$ solutions (see Panel (h)) shows systematic behavior. This trend in the range bias can be explained from the contribution of the relativistic deflection of light by the Sun (a function of \gls{PPN} parameter $\gamma$, \cite{shapiro64}) in the light time computations. Explicitly this effect was not absorbed during the computation of range bias and it becomes important to examine this effect when constructing planetary ephemerides. 

We, therefore, reconstruct the planetary ephemerides using the range bias obtained from SOL$_{\beta\gamma\neq1}$ and the \gls{PPN} parameters $\beta$-1 = $\gamma$-1 = 1$\times$10$^{-4}$. The newly estimated postfit range bias is then compared with the range bias (prefit) corresponding to SOL$_{\beta\gamma\neq1}$. This investigation shows that, the postfit residuals are modified by $\sim$6$\%$ for \gls{MGR} and $\sim$1$\%$ for Mariner 10 with respect to prefit residuals. This modification in the residuals is negligible compared to a $\sim$65$\%$ of change with respect to reference residuals obtained from INPOP13a. As a result, the supplementary contributions in the range bias due to the relativistic deflection between \gls{DSN} station and \gls{MGR} did not bring any significant change in the planetary ephemerides construction.


\section{Conclusions}
\label{dis}
We analyzed one and half years of radioscience data of the \gls{MGR} spacecraft using orbit determination software GINS. An accurate orbit of \gls{MGR} was then constructed with the typical range of Doppler, and two-way range residuals of about 4.8$\pm$2.2 mHz ($\sim$0.09$\pm$0.04 mm/s), and 1.9$\pm$1.4 m. Such accuracies are comparable to those in \cite{Smith12,Stanbridge11,Srinivasan07}. Range measurements obtained by the MESSENGER spacecraft during its mapping period were then used to construct improved planetary ephemerides called INPOP13a. This ephemeris showed an accuracy of about -0.4$\pm$8.4 m in the Mercury-Earth geometric distances, which is two orders of improvement compared to DE423 and INPOP10e, and one order compared to the latest DE430.

Such high precision Mercury ephemeris allowed us to perform one of the best \gls{GR} tests of \gls{PPN}-formalism. To determine the acceptable intervals of the violation of \gls{GR} through the \gls{PPN} parameters ($\beta$, $\gamma$), small variations of these two parameters near unity were imposed in the construction of alternative planetary ephemerides fitted over the whole data sets. The percentage difference between these ephemerides to INPOP13a are then used to defined the interval of \gls{PPN} parameters $\beta$ and $\gamma$.

As expected, our estimations of PPN parameters are more stringent than previous results. We considered the 5$\%$, 10 $\%$ and 25$\%$ of changes in the postfit residuals. That the \gls{PPN} intervals correspond to these changes is compatible with \gls{GR} with an uncertainty at least ten times smaller than our previous results with INPOP10a. Moreover, one of the best estimation of parameter $\gamma$ has so far been estimated from the Cassini observations by \cite{Bertotti03}, which is compatible with our 25$\%$ estimation.

{ 
To further the accuracy of the \gls{PPN} parameters improve, and to confirm the results given in Table \ref{paramfitc}, one needs to analysis the radioscience data of the future Bepi-Colombo mission.}

\section{Acknowledgments}
We are very thankful to the CNES and Region Franche-Comte, who gave us financial support. Part of this work useds GINS software, so we would like to acknowledge the CNES for providing access to this software. A. K. Verma is thankful to P. Rosenbatt and S. Le Maistre for fruitful discussions. We are also thankful to G.Esposito-Farese for his constructive remarks and comments. 
  
\glsaddall
\printglossaries

\bibliographystyle{aa}
\bibliography{reference2.bib}
\end{document}